\documentclass[twocolumn,nodate,showpacs,preprintnumbers,nofootinbib,amsmath,amssymb,aps,prd]{revtex4}

\usepackage{color}
\usepackage{graphicx}
\usepackage{amsmath,amssymb}

\usepackage{float}

\usepackage[colorinlistoftodos]{todonotes}
\usepackage{epstopdf}

\def\be{\nopagebreak[3]\begin{equation}}
\def\ee{\end{equation}}
\def\ba{\nopagebreak[3]\begin{eqnarray}}
\def\ea{\end{eqnarray}}

\begin{document}

\title{Loop quantum cosmology, non-Gaussianity, and CMB power asymmetry}

\author{Ivan Agullo} \email{agullo@lsu.edu}
\affiliation{Department of Physics and Astronomy, Louisiana State University, Baton Rouge, LA 70803}

\begin{abstract}

We argue that the anomalous power asymmetry observed in the cosmic microwave background (CMB) may have originated in a cosmic bounce preceding inflation.  
In loop quantum cosmology (LQC) the big bang singularity is generically replaced by a bounce due to quantum gravitational effects. 
We compute the spectrum of inflationary non-Gaussianity and show that strong correlation between observable scales and modes with  longer (super-horizon) wavelength  arise as a consequence of the evolution of perturbations across the LQC bounce.  These correlations are strongly scale dependent and induce a dipole-dominated modulation on large angular scales in the CMB, in agreement with observations.

\end{abstract}


\maketitle

\section{Introduction}
\label{s1}

The {\em Planck} team has reported evidence of anomalous features in the large-scale cosmic microwave background (CMB) which point towards violations  of statistical isotropy  \cite{Planck2013Isotropy}. Some of these anomalies were already observed by WMAP, and have been now  re-confirmed \cite{Planck2015Isotropy}, hence reducing the possibility of instrumental origin or systematics.  
Although the associated statistical significances are still inconclusive, the intriguing possibility that these features  are traces of fundamental physics beyond the simplest inflationary models have  attracted considerable attention in the theoretical community.

Among the different anomalies detected, the {\em power asymmetry}---unequal power in different regions of the sky---has received significant attention, and it is also the main target of this paper. This asymmetry was first modeled by adding a modulating factor to an otherwise  statistically isotropic temperature distribution \cite{gordonetal}
\be \label{mod} \delta T(\hat{n})=\big(1+A(\hat{n})\big)\,  \delta T_{\rm iso}(\hat n) \, \ee
where $A(\hat{n})=\sum_{LM} A_{LM}\, Y_{LM}(\hat{n})$ is the modulating function.   {\em Planck} has reported no evidence for $A_{LM}$ different from zero for $L \geq 2$, but a statistically significant (about $3\sigma$) $L=1$, {\em dipolar} modulation has been found \cite{Planck2013Isotropy}. Furthermore, the effect  of this dipole on the angular power spectrum $C_{\ell}$  has  been detected {\em only} for low multipoles with $\ell < 64$. More precisely,  after separating the $\ell$-range into bins of size $\Delta\ell=64$, only the first bin shows a non-vanishing signal, with average amplitude $A_{L=1}=0.07\pm 0.02$. Observations are compatible with $A_{L=1}=0$ for larger values of $\ell$. This scale dependence implies that the simple parametrization (\ref{mod}) is insufficient to account for the observed modulation. The theoretical challenge  is therefore to find a mechanism able to produce a dipolar modulation present only on large scales, with negligible contribution to the quadrupole, octupole, etc, and furthermore respecting the existing  constraints on the CMB: a remarkably Gaussian, almost-scale invariant spectrum of adiabatic perturbations.  Not surprisingly,  it has been  difficult to find a completely satisfactory model \cite{pesky}.

One of the first and most compelling ideas to generate such a power asymmetry was introduced by Erickcek, Carroll, and Kamionkowski in \cite{eck}. It relies on the fact that the presence of a very long wavelength, super-Hubble perturbation of a curvaton field will induce a dipole modulation in the observed spectrum, provided its wavelength is still short enough that its amplitude shows a gradient across the sky. This long mode could originate, for instance, as a remnant of the pre-inflationary epoch. This idea has been implemented by several authors and improved in different  directions  (see e.g. \cite{djkc,lc,lyth}). 
 
 Another interesting proposal, introduced in \cite{halo-bias3}, works instead in a single-field inflationary model, and generates the power modulation through a non-Gaussian {\em coupling} between observable scales in the CMB and even larger, super-Hubble scales. The reason why non-Gaussianity  can induce anisotropies in the observed spectrum, even if the underlying statistics is  isotropic, is simply because a typical realization  looks significantly more anisotropic if the underlying distribution is non-Gaussian. Or in other words,  because observable modes couple to the {\em  particular  realization} of  the long wavelength modes in our Universe, which is generically anisotropic.  This model, therefore, requires a mechanism to generate large correlations between very different scales. The consistency relation proposed in \cite{cz} tell us that it is difficult to find a realistic  single field model producing that type of non-Gaussianity.

In this work we introduce a scenario which has features in common with both of the previous  ideas.  We consider a single field inflationary model preceded by a bounce described by loop quantum cosmology (LQC). This scenario has been analyzed in great  detail, both for the background space-time and for perturbations (see \cite{asrev,agullo-corichi,ashtekar-barrau, barraureview} for reviews). In short, the evolution of perturbations across the bounce excites quanta out of an initial vacuum, and as a consequence the onset of inflation is reached in an excited state, rather than in the Bunch-Davies vacuum. The presence of those perturbations,  remnants of the pre-inflationary phase, have an important impact on the non-Gaussianity generated during inflation. We compute those non-Gaussianity and the modulation they  produce on observable scales, and show that the observed power asymmetry can have an origin on the quantum bounce preceding inflation, while still respecting all observational constraints. 
We set $c$=1 but keep $G$ and
$\hbar$ explicitly. Numerical values are given in Planck units.

\section{LQC and the power spectrum}

In LQC, the mean effective space-time geometry is  described by  equations which incorporate the leading quantum corrections to general relativity. For instance, the Friedmann equation reads \cite{vt,asrev}
\be \label{Feq} H^2=\frac{8\pi G}{3}\rho \,  \left(1-\frac{\rho}{\rho_{\rm max}}\right)\, ,\ee
where $H$ is the Hubble rate, $\rho$ the energy density, and $\rho_{\max}\approx 0.4 \rho_{Pl}$ its {\em upper bound}, which is a fraction of the Planck energy density $\rho_{Pl}$. In this paper the matter sector is assumed to be a single scalar field with mass $m$. More complicated potentials can be incorporated. However,  the effects we are seeking  have a quantum gravity origin, and the form inflaton potential produces sub-leading contributions.  At low energies $\rho\ll \rho_{\rm max}$, Eq. (\ref{Feq}) becomes indistinguishable from the general relativistic Friedmann equation. However, close to the Planck scale quantum gravitational effects  break the linear relation between $H^2$ and $\rho$, and bring $H$ to zero while $\rho$ attaints its maximum value. This is a quantum bounce that `bridges' a contracting and an expanding phase of the universe.  

The theory of cosmological perturbations in LQC that we use in this paper was developed in \cite{aan},  and has been  used in \cite{aan,am} to study the evolution of perturbations across the bounce through the end of inflation, and to compute the primordial power spectrum of tensor and scalar perturbations. This scenario is based on first principles, and trans-Planckian issues can be addressed squarely. The  free parameters relevant for phenomenological studies are the value of the inflaton field at some arbitrary reference time,  e.g.\  the bounce, $\phi(t_B):=\phi_B$, its mass $m$, and the initial state of perturbations at some initial time. Refs. \cite{aan,am} have explored the predictions for the power spectrum across the parameter space and contrasted the results with observations.

From the viewpoint of inflation, the effect of the preceding bounce translates to an excited  state  $|\beta\rangle$ for perturbations at the onset of slow-roll. Therefore, the LQC pre-inflationary evolution can be conveniently encoded in the Bogoliubov coefficients $\alpha_k$ and $\beta_k$ relating $|\beta\rangle$ and the Bunch-Davies vacuum,   i.e.\  relating the mode functions of curvature perturbations $\mathcal{R}_{k}(t)$ that result from the pre-inflationary evolution, and the Bunch-Davies modes $\mathcal{R}^{\rm BD}_{k}(t)$. These coefficients can computed numerically for a choice of the free parameters. For $\phi_B=1.22$ and $m=1.10\times 10^{-6}$, both in Planck units, and ``Minkowski-like'' vacuum initial condition\footnote{As explained in \cite{aan}, although Minkowski-like initial conditions do not produce a state with the desired ultraviolet behavior---i.e.\ a Hadamard or fourth adiabatic order state---one can always modify the initial data for sufficiently  ultraviolet modes to make the state ultraviolet-regular. Since such modification do not affect observable predictions, we do not describe  the details  in this paper, which can be found in \cite{aan,ana}} for perturbations at one Planck second before the bounce, we obtain that the average number of quanta present at the onset of inflation, given by $|\beta_k|^2$, is approximately $10^{-3}$ for the reference mode $k_{\star}$ which today corresponds to $0.002\,{\rm Mpc}^{-1}$. For the longest wavelength mode we can  directly measure, $k_{\rm min}\approx k_{\star}/8.9$, we obtain $|\beta_{k_{\rm min}}|^2= 1.2$, and  $|\beta_k|^2\sim 1/k$ for lower values of $k$ \cite{am}. Other choices of initial data give similar results (see \cite{aan,am} for details).  

\begin{figure}[h]
\begin{center}
\includegraphics[width=0.45\textwidth,angle=0]{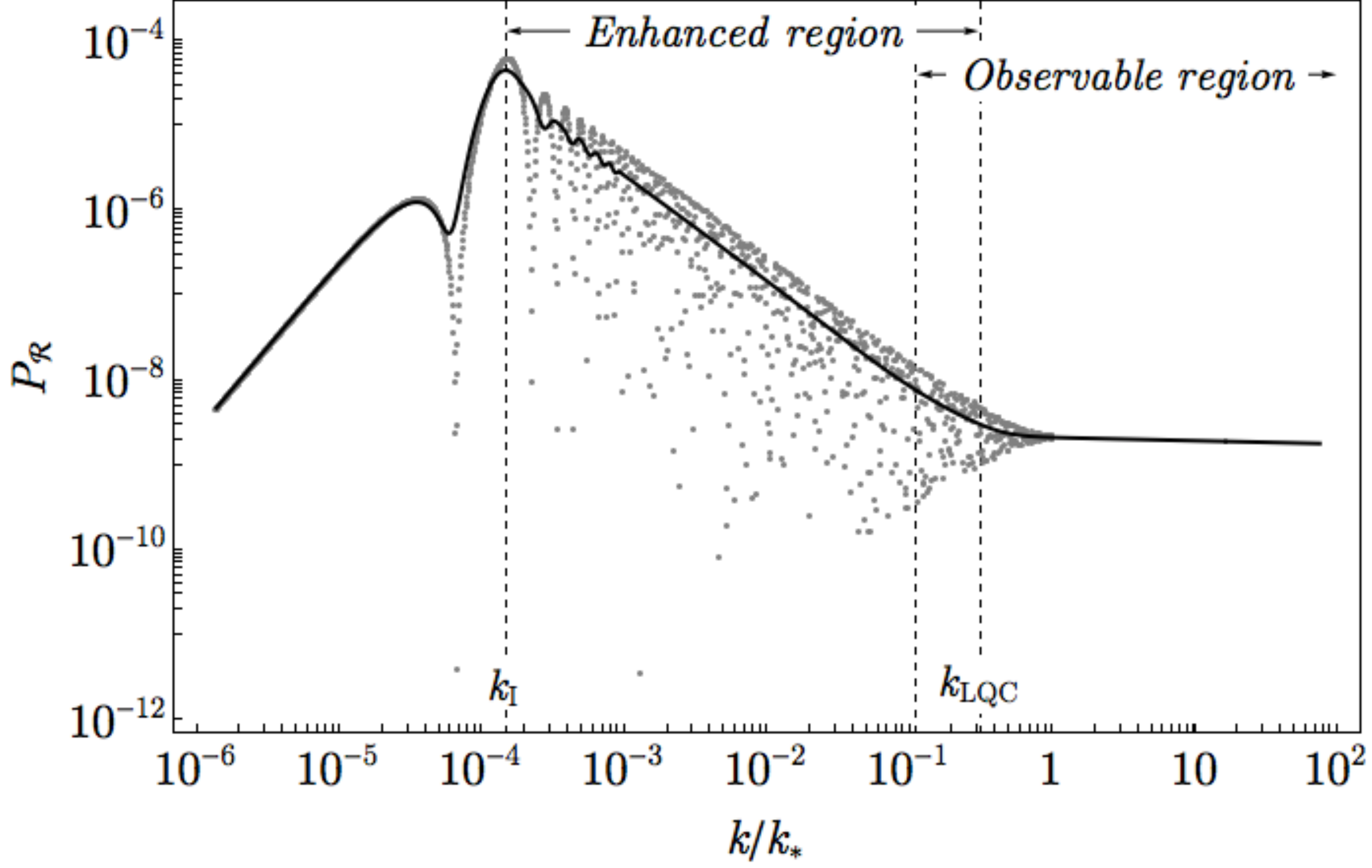} 
\caption{Scalar power spectrum at the end of inflation for $\phi_B=1.22$ and $m=1.10\times 10^{-6}$, and ``Minkowski-like'' vacuum initial conditions at one Planck second before the bounce \cite{am}. Gray points show the numerical result for individual modes. The spectrum is rapidly oscillatory, and its average is shown in black. The spectrum is amplified for low wave-numbers $k_I<k<k_{LQC}$, but the enhancement is only significant for the very low multipoles in the CMB, for which the observational error bars are large because cosmic variance. Therefore, this power spectrum is compatible with current observations.  \label{fig1}}
\end{center}
\end{figure}

The inflationary scalar power spectrum in the presence of an excited state $|\beta\rangle$ is given by $P_{\mathcal{R}}(k)=P_{\mathcal{R}}^{\rm BD}(k)\, |\alpha_k+\beta_k|^2$, where $P_{\mathcal{R}}^{\rm BD}(k)$ is the Bunch-Davies result. Fig. \ref{fig1} shows the numerically computed $P_{\mathcal{R}}(k)$ for the choice of parameters mentioned above. Two energy scales play an important role in the power spectrum. First, LQC introduces a {\em new energy scale} $k_{LQC}/a(t_B):= \sqrt{R_B/6}\approx 3.21$, that is directly related to the space-time scalar curvature  at the bounce $R_B=48\pi \rho_{\rm max}\approx 62$.  A second  scale is provided by the value of curvature at the onset of accelerated expansion $k_{I}/a(t_{I}):= \sqrt{R_I/6}\approx  10^{-5}$, which occurs at time $t_I$---notice that $t_I$ is not the onset of {\em slow-roll}, which happens at later times. Modes $k>k_{LQC}$ are `inside' the curvature radius---i.e.\ its wavelength is shorter than the curvature radius---during the bounce, and exit during the slow-roll era. These perturbations reach the onset of slow-roll in the Bunch-Davies vacuum, and their power spectrum is indistinguishable from  standard results. Modes $k_I<k<k_{LQC}$ exit the curvature radius soon before the bounce and re-enter after it, to exit again during slow-roll. The crossing process around the bounce time amplifies the amplitude of those modes, which then reach inflation in an excited states. Finally,  modes $k<k_I$ are outside the Hubble radius during the entire evolution, even at the onset of inflation, and their spectrum is significantly suppressed. The reference mode $k_{\star}$, that today corresponds to $0.002\,{\rm Mpc}^{-1}$, is approximately one third of $k_{LQC}$, $k_{\star}=0.36\, k_{LQC}$. Consequently, the LQC corrections to the observable power spectrum only appear for the largest angular scales in the sky, that correspond to multiples $\ell\lesssim 30$, and are only significant for the lowest multiples. Furthermore, the LQC corrections significantly increases the power for modes with wavelengths larger than the Hubble radius today. This power enhancement conveniently reaches a maximum  around $k_I$, at which perturbation theory is still well under control.
\section{Back-reaction on the inflationary geometry}

An important question is whether the energy in the quanta present at the onset of slow-roll produces significant back-reaction  on the inflationary space-time geometry which cannot be neglected. The energy density on the perturbations can be computed and compared to the vacuum result. The {\em difference}  in energy density between  the state $|\beta\rangle$ and the Bunch-Davies vacuum can be written as
\ba \label{deltarho} \Delta\rho (t)=\frac{\hbar \, \epsilon}{ 4\pi G} \int \frac{d^3k}{(2\pi)^3} \Big\{ |\beta_k|^2 \big(|\dot {\mathcal{R}}_k^{BD}|^2+\frac{k^2}{a^2} |\mathcal{R}_k^{BD}|^2\big) \nonumber \\+\, {\rm Re} \Big[\alpha_k \beta_k^{\star}\, [ (\dot{\mathcal{R}}_k^{BD})^2+\frac{k^2}{a^2} (\mathcal{R}_k^{BD})^2]\Big]\Big\} \, , \ea
where the integral in $k$ is extended from $-\infty$ to $\infty$.
It is well known that the difference of expectation values of the energy-momentum tensor between  states that are at least of fourth adiabatic order is always finite, hence no renormalization is required in (\ref{deltarho}). Numerical evaluation shows that the ratio of $\Delta\rho$ with the background energy density $\rho_0$, is $\Delta\rho/\rho_{\rm 0} \approx 10^{-3}$ at the onset of slow-roll, and  decreases  exponentially  with cosmic time. Therefore, since the energy density in the Bunch-Davies vacuum is  known to be negligible small, the back-reaction of the state $|\beta\rangle$ on the inflationary geometry  can be neglected.

\section{The bispectrum}
At leading order in perturbation theory, non-Gaussianity is characterized by the bispectrum $B_{\mathcal{R}}(k_1,k_2,k_3)$, defined in terms of the three-point function:

\be \langle {\mathcal{R}}_{\vec{k}_1}{\mathcal{R}}_{\vec{k}_2}{\mathcal{R}}_{\vec{k}_3}\rangle =(2\pi)^3\delta^3(\vec{k}_1+\vec{k}_2+\vec{k}_3)\;B_{{\mathcal{R}}}(\vec{k}_1,\vec{k}_2,\vec{k}_3) \, . \nonumber \ee
The  non-Gaussianity generated from excited states in inflation has been analyzed by several authors (see e.g. \cite{NG, ap, ganc, as}).  It was pointed out in  \cite{ap} that the main characteristic of the  associated bispectrum  is an enhancement in `squeezed' configurations  which involve very different scales, $k_1\approx k_2\gg k_3$.  These non-Gaussian correlations originate from quantum interactions between  particles that are present at the onset of slow-roll. In realistic models, including the scenario presented in this paper,  the conclusions of the consistency relation proposed in \cite{cz} are still satisfied in the limit $k_1/k_3\to 0$, but  nevertheless there are  important effects for small, but finite, values of $k_1/k_3$. 

\begin{figure}[h]
\begin{center}
\includegraphics[width=0.48\textwidth,angle=0]{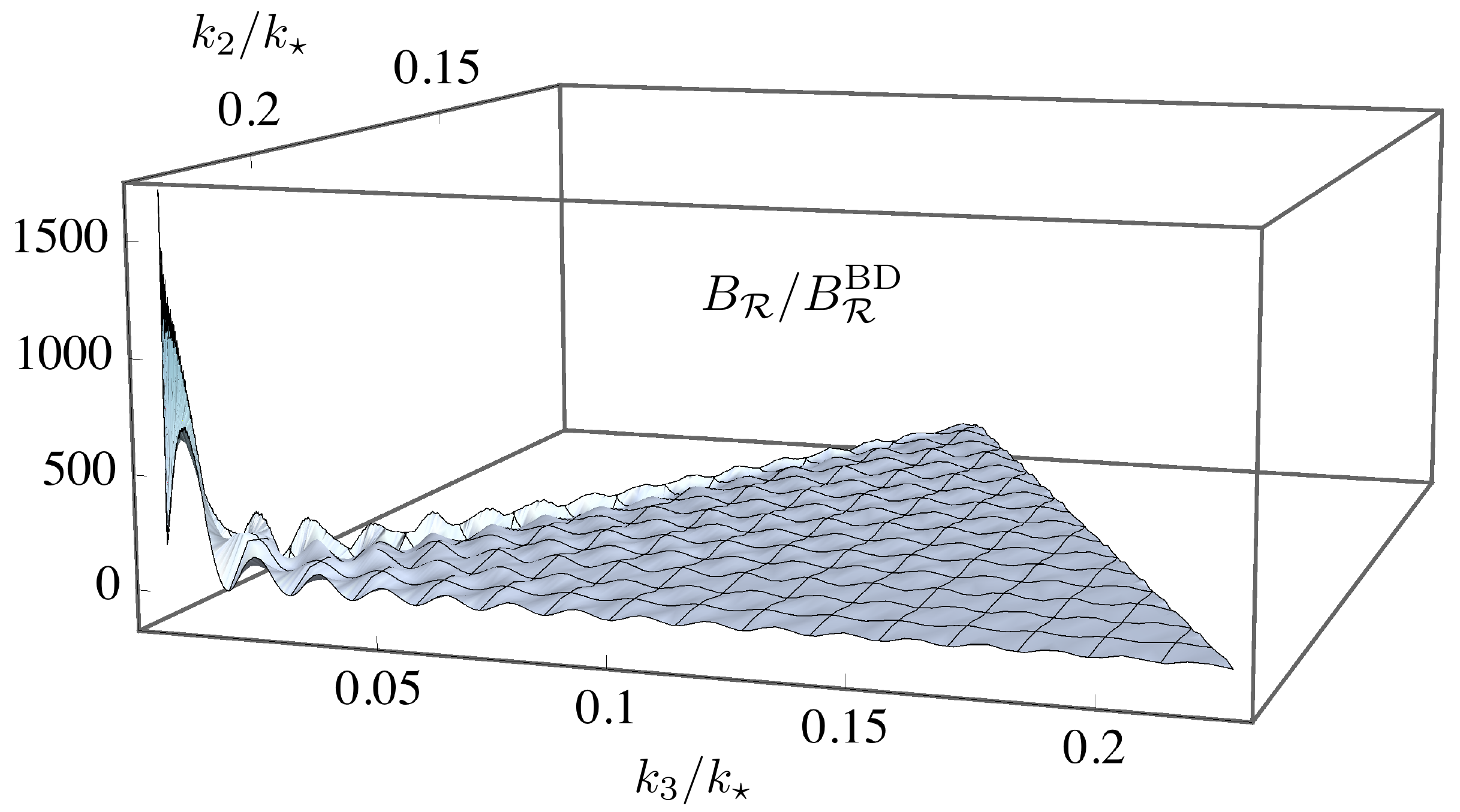} 
\caption{Ratio of the inflationary bispectrum for the excited initial state  arising from the LQC pre-inflationary evolution versus the Bunch-Davies bispectrum, as a function of $k_2$ and $k_3$ for a fixed $k_1=0.22 k_{\star}$. The plot shows the range of $k_2$ and $k_3$ allowed by the triangle condition $\vec{k}_1+\vec{k}_2+\vec{k}_3=0$. The bispectrum $B_{\mathcal{R}}$ is highly scale dependent and shows  a prominent enhancement for squeezed configurations $k_3\ll k_2\approx k_1$. \label{fig2}}
\end{center}
\end{figure}

The full expression for the bispectrum $B_{\mathcal{R}}(k_1,k_2,k_3)$  as a function of the state $|\beta\rangle$ was obtained in \cite{ap,ganc, as}, and is given in the appendix. We have used those results  to numerically computed  $B_{\mathcal{R}}(k_1,k_2,k_3)$ at the end of inflation. The main difference with previous analysis is that the state $|\beta\rangle$ is now  {\em computed}  from a  quantum gravity framework, rather than postulated. The resulting bispectrum, therefore, carries information about the pre-inflationary evolution.  Those effects are more clearly displayed by plotting the ratio with the standard Bunch-Davies bispectrum $B_{\mathcal{R}}^{\rm BD}(k_1,k_2,k_3)$ (see Fig.\ \ref{fig2}). As expected, the Bispectrum is strongly peaked in squeezed triangles, where the ratio   $B_{\mathcal{R}}/B_{\mathcal{R}}^{\rm BD}$ shows a growth proportional to $\frac{k_{LQC}^2}{k_1 k_3}$\, ---i.e.\ proportional to product of the number of quanta in the modes $k_1$ and $k_3$, as expected for an enhancement originated from particle interactions. As for the power spectrum, the enhancement {\em only} appears for modes $k_I\lesssim k\lesssim k_{LQC}$. Therefore, there are strong correlations between the largest wavelengths we can directly observe ($k\approx k_{LQC}$) and super-Hubble modes with  $k$ values in the range $k_I\lesssim k\lesssim k_{LQC}$. The bispectrum becomes negligibly small when the three momenta are in observable scales, hence respecting current constraints on non-Gaussianity. This scale dependence of the  bispectrum will play an important role in the next section.

Recall that the three modes involved in bispectrum must form a triangle, so it is really a function of two momenta, e.g.\ $B_{\mathcal{R}}(\vec{k}_1,\vec{k}_3)$. The shape of the bispectrum, shown in Fig. \ref{fig2}, can be understood by writing its dominant contribution  in the squeezed limit $x:=k_3/k_1\ll 1$:
\ba  \label{bisp}  B_{\mathcal{R}}\approx  4\epsilon \, \Delta_{\mathcal{R}}(k_1)\Delta_{\mathcal{R}}(k_3)\, \times \, {\rm Re} \Big [f_t\frac{1-e^{i  \tilde{k}_t \eta_0}}{1+(1+\mu )\, x}+\nonumber \\   f_1 \frac{1-e^{i \tilde{k}_1\eta_0}}{(1+\mu )\, x}+f_2\frac{1-e^{i \tilde{k}_2\eta_0}}{(1-\mu)\, x}+  f_3\frac{1-e^{i \tilde{k}_3\eta_0}}{1+(-1+\mu )\, x}\Big]\ea
where $\mu=\hat k_1\cdot\hat k_3$, $k_t=k_1+k_2+k_3$, $\tilde{k}_i=k_t-2 k_i$, and $\Delta_{\mathcal{R}}(k)=\hbar \frac{2\pi^2 }{k^3} \left(\frac{H}{\dot \phi}\right)^2\left(\frac{H}{2\pi}\right)^2$ evaluated at Hubble-crossing during inflation. The functions $f_i$ contain the information of the state $|\beta\rangle$ and are defined in the appendix.  The parameter $\eta_0$ is the value of the conformal time at the onset of slow-roll.  For the example $\phi_B=1.22$ and $m=1.1\times 10^{-6}$ we obtain $\eta_0 k_{\star}\approx 10^{3}$, so all observable modes are deeply inside the Hubble radius at the onset of slow-roll.

The largest  contribution to (\ref{bisp}) comes  from very squeezed configurations for which $\tilde{k}_3\eta_0\ll 1$. In that limit $B_{\mathcal{R}}$ is dominated by the terms proportional to $f_t$ and $f_3$. This contrasts with the example considered  in \cite{ap,as} of a scale invariant excited state $|\beta\rangle$,  and where additionally the analysis was restricted to observable scales for which $\eta_0 k_i\gg 1$ for all $i$. In that case the leading contributions come from the terms proportional to  $f_1$ and $f_2$.

\begin{figure*}
\begin{center}
\includegraphics[width=\textwidth,angle=0]{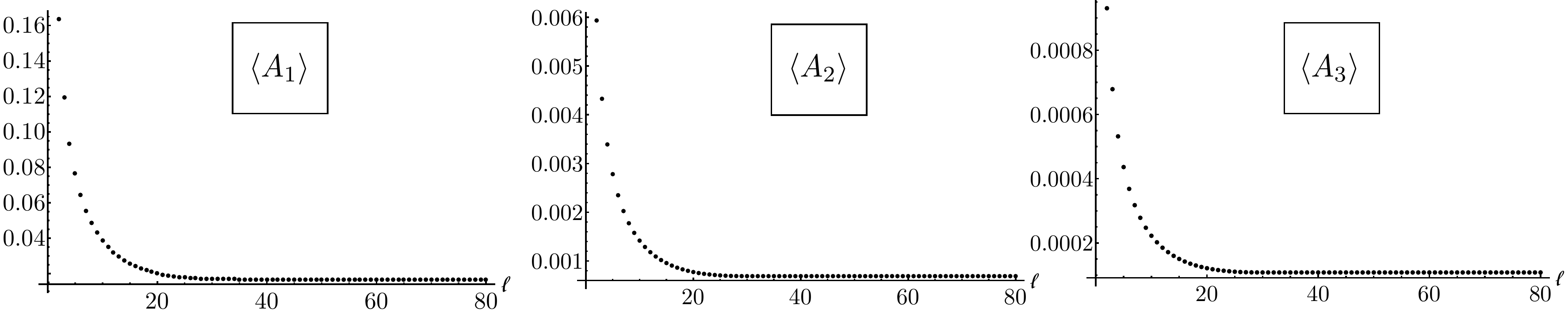} 
\caption{Amplitude  $\langle A_L(\ell)\rangle$ as a function of $\ell$ for L=1,2,3, for parameter values $\phi_B=1.22$ and $m=1.10\times 10^{-6}$, and ``Minkowski-like'' vacuum initial conditions at one Planck second before the bounce.}
\label{fig3}
\end{center}
\end{figure*}

It is important to note that the non-Gaussianity computed here are generated during slow-roll inflation. Extra contributions to the bispectrum will certainly arise from the evolution across the bounce. However, since the state of perturbations is very close to vacuum at the bounce time, those contributions are expected to be   subdominant. This expectation is indeed borne out  in explicit computations in bouncing models \cite{brandenberger-cai, Sreenath}, where it is shown that the non-Gaussianity generated across the bounce  are much smaller than the bispectrum shown in Fig. \ref{fig2}, particularly for squeezed triangles, which are the relevant configurations for this paper. It is therefore reasonable to assume that the leading contributions to the bispectrum in squeezed configurations are dominated by inflationary non-Gaussianity.

\section{Non-Gaussian Modulation}
We follow some  of the  ideas presented in Ref. \cite{halo-bias3} to compute the CMB power modulation arising from  coupling with super-Hubble scales. The statistics of the classical Bardeen potential $\Phi$ for observable scales  is modified in the presence of a given long wavelength perturbation  $\Phi(\vec{k}_l)$ if the spectrum is non-Gaussian. Concretely, the two-point function acquires off-diagonal contributions  of the form \cite{lewis, jk}
\be \label{2pf} \langle \Phi_{\vec{k}} \Phi_{\vec{k}'}\rangle = P_{\Phi}(k)\, \left[ (2\pi)^3 \delta(\vec{k}+\vec{k}') +G(\vec{k},\vec{k}_l)\, \Phi(\vec{k}_l)\right] \, \ee
where $\vec{k}_l=-(\vec{k}+\vec{k}')$ (i.e.\  it  closes a triangle with $\vec{k}$ and $\vec{k}'$),  $P_{\Phi}(k)$ is the $\Phi$-power spectrum, and we take $G(\vec{k},\vec{k}_l)=\frac{5}{3}B_{\mathcal{R}}(\vec{k},\vec{k}_l)[\Delta_{\mathcal{R}}(k)\Delta_{\mathcal{R}}(k_l)]^{-1}$. For bispectra  peaked in squeezed configurations the leading contributions to  $G(\vec{k},\vec{k}_l)$ come from the regime $k_l\ll k$, for which $k\approx k'$. Therefore, the correlation function is close to be diagonal. 
 The off-diagonal terms break both homogeneity and isotropy.  These terms vanish if we average over $\Phi_{\vec{k}_l}$, as it must be, since our model respects these symmetries at the fundamental level.  But in the particular realization of $\Phi_{\vec{k}_l}$ chosen  by our Universe, the modulation term  may be important, and may  produce  deviations from homogeneity and isotropy in the CMB much larger than would be expected from a typical realization of a  Gaussian spectrum.

The off-diagonal terms in (\ref{2pf}) source analogous terms in the covariance matrix of temperature  spherical harmonic coefficients 
\be \nonumber \langle a_{\ell m}a^{\star}_{\ell' m'}\rangle= \delta_{\ell\ell'}\delta_{mm'}C_{\ell}+\sum_{LM} A_{LM}\,  \mathcal{G}_{-mm'M}^{\ell\ell' L}\,  (C_{\ell}+C_{\ell'})\, ,\ee
where $\mathcal{G}_{m_1m_2m_3}^{\ell_1\ell_2 \ell_3}$ is a Wigner 3-$j$ symbol, and $C_{\ell}$ is the standard angular power spectrum. The momentum dependence of the kernel $G(\vec{k},\vec{k}_l)$ factorizes for the dominant contributions in a power law for $k$ (except for typical oscillations which are unimportant for our computations and can be averaged). The kernel can be expanded in Legendre Polynomials 
$$G(\vec{k},\vec{k}_l)=\sum_{L} g_L(k_l)\, \left(\frac{k_{\star}}{k}\right)^{\alpha_L} P_L(\mu)\, ,$$
where $\mu=\hat k\cdot \hat k_l$, and $g_L(k_l)$ encodes the $k_l$ dependence. Then, the mean value of the modulation amplitude, averaged over $M$, for a typical realization of the long wavelength modes  can be computed as (see \cite{halo-bias3} for further details of the computation)
\be  \langle A_L(\ell)\rangle=\frac{C_{\ell}(-\alpha_L)}{2 C_{\ell}(0)}\left[\int \frac{dk_l\, k_l^2}{(2\pi)^3}\, |g_L(k_l)|^2 P_{\Phi}(k_l) \right]^{1/2}\, ,\ee
where the coefficients $C_{\ell}(\alpha_L)$ have been defined as the CMB  temperature power spectrum computed by replacing $n_s\to n_s+\alpha_L$. Therefore, if $\alpha_L=0$ for all $L$, i.e.\ if  $G(\vec{k},\vec{k}_l)$ does not depend on $k$, the modulation amplitude  is $\ell$-independent.

The value of $\langle A_L(\ell)\rangle$ for different multipoles $L$ can be computed numerically from the bispectrum shown in Fig. (\ref{fig2}), and some of its properties  can be qualitative understood already from equation (\ref{bisp}). The leading contribution comes from the terms proportional to $f_t$ and $f_3$, which are dominated by a {\em dipole} $L=1$. The terms proportional to $f_1$ and $f_2$ contribute to a subdominant quadrupole. The octopole $L=3$  is subdominant compared to the quadrupole, and this hierarchy continues for higher multipoles. Since the factors $f_i$ in (\ref{bisp}) scale approximately as $(k_l \,k)^{-1}$ for every $i$, we have $\alpha_L\approx 1$ for all $L$, and therefore the amplitude $A(\ell)$ is expected to {\em decrease} with $\ell$. 

Figure \ref{fig3} shows $\langle A_L(\ell)\rangle$ for $L=1,2,3$ as a function of $\ell$. The amplitudes $\langle A_L(\ell)\rangle$ are all scale dependent, and such that $\langle A_1\rangle\gtrsim  10 \, \langle A_2\rangle \gtrsim 10\,  \langle A_3\rangle$. The average value of $\langle A_1\rangle$ for $\ell\lesssim 64 $ is in agreement with the observed value $A_{1}^{\rm obs}=0.07\pm0.02$. Other values of our free parameters $\phi_B$ and $m$, and of the initial conditions for perturbations, can decrease or increase the value of  $\langle A_1\rangle$. For instance, we find that choosing vacuum initial data for perturbations far into the past of the bounce, significantly decreases the amplitudes of all multipoles.

There has been some debate based on symmetry arguments about whether a statistically homogeneous and isotropic bispectrum can generate a dipole modulation. Our computation is an example where the answer is in the affirmative. Furthermore,  as shown in \cite{baumann}, a careful analysis reveals  that symmetry arguments restrict   $A_{L+1}$ to be suppressed with respect to $A_L$ for even $L$. The hierarchy of multipoles we find  here,  $\langle A_0\rangle > \langle A_1\rangle>\langle A_2\rangle>...$., is therefore in agreement with those restrictions.

\section{Discussion}

We have presented an inflationary scenario in which perturbations start the slow-roll phase in an excited state, rather than the Bunch-Davies vacuum. This state arises from the pre-inflationary evolution provided by loop quantum cosmology, in which the big bang singularity is replaced by a bounce. Two new scales appear in the problem, $k_I$ and $k_{LQC}$, related to the onset of the exponential expansion and the bounce, respectively. 
The number density of quanta at the onset of slow-roll is significant only for the range  $k_I\lesssim k\lesssim k_{LQC}$. During inflation these excitations induce non-Gaussian correlations  which we have computed. To the best of our knowledge, this is the first computation of non-Gaussianity in LQC. The result  is compatible with existing observational constraints. Furthermore, large correlations arise between the longest modes we can observe, with $k\approx k_{LQC}$, and super-Hubble modes with $k\gtrsim k_I$.  
We have shown that  those non-Gaussian correlations, which involve super-Hubble modes, are able to modify the observed power spectrum at large scales, inducing correlations  between  CMB angular multipoles $\ell\lesssim 30$ that differ in $\Delta\ell\approx 1$. These correlations are strongly scale dependent, and produce a power asymmetry in agreement with observations. 

Other observed anomalies at large scales \cite{Planck2013Isotropy}---parity violation, power suppression for a bunch multipoles around $\ell=20$, multipole alignments, etc.---seem also to require correlations between low multipoles qualitatively similar to the ones obtained in this paper. Whether the agreement is also quantitative will be analyzed in future work.

We emphasize that the fact that the  power spectrum shown in Fig.\ \ref{fig1}  is amplified for low wave-numbers $k_I<k<k_{LQC}$, is not necessarily in conflict with observations since: i)  the enhancement is only significant for the very low multipoles in the CMB, for which the observational error bars are large because of cosmic variance; ii) the observed power suppression is only significant for a few multipoles around $\ell=20$, indicating that the effect  is more likely to originate from correlations between multipoles, rather than a suppression of the primordial two-point function.

Some of  the features appearing in our scenario---remnants, large amplitude perturbations associated to an infrared scale $k_I$, correlations with super-Hubble modes, etc.---have been identified in previous phenomenological analysis (see e.g.\ \cite{eck, halo-bias3}) as ingredients needed to account for some of the observed  anomalies at large scales. Here, these features arise from a concrete quantum gravity proposal based on first principles. Therefore, our results   provide further motivation to  consider the observed anomalies as real physical features, which have origin beyond the simplest inflationary models, rather than statistical flukes or instrumental noise. Future work will be addressed to provide further robustness to the model introduced here and to extend its quantitative predictions. \\

\section*{Acknowledgments}

We thank  A. Ashtekar, D. Baumann, B. Gupt, N. Morris, and  S. Shandera for useful discussions. This work is supported by the NSF Grant No. PHY-1403943.
\begin{appendix}
\section{Inflationary bispectrum from an excited state}
\label{a1}

The expression for the scalar bispectrum generated during slow-roll inflation when the state of perturbations $|\beta\rangle$ is given by a Bogoliubov  transformation of the Bunch-Davies vacuum, with coefficient $\alpha_k$ and $\beta_k$, is given by \cite{NG, ap, ganc, as}

\begin{widetext}
\ba \label{fullbisp}
& &B_{\mathcal{R}}(k_1,k_2,k_3) = \Delta_{\mathcal{R}}(k_1)\Delta_{\mathcal{R}}(k_2)  \Big\{ \frac{1}{2} \left(3 \epsilon-2\eta +\epsilon \,  \frac{k_1^2+k_2^2}{k_3^2} \right) + \\\nonumber
&+&\, 4 \epsilon \, \frac{k_1^2k_2^2}{k_3^3} \ \textrm{Re}\left[f_t\frac{1-e^{ik_t\eta_0}}{k_t}+f_{1}\frac{1-e^{i\tilde{k}_1\eta_0}}{\tilde{k}_1}+f_{2}\frac{1-e^{i\tilde{k}_2\eta_0}}{\tilde{k}_2}+f_{3}\frac{1-e^{i\tilde{k}_3\eta_0}}{\tilde{k}_3} \right]  \Big\}  +\textrm{2 cyclic permut. $k_1\to k_2\to k_3$}\, ,\ea
\end{widetext}
where $k_t=k_1+k_2+k_3$, $\tilde{k}_i=k_t-2 k_i$, and the parameter $\eta_0$ is the value of the conformal time at the onset of slow-roll.
The functions $f_i$ contain the information of the state $|\beta\rangle$:
 $$f_t=\big[ 1+ F(k_1)\, (1+F(k_2))+{\rm cyclic  \, perm.}\big]\, ,$$
 $$f_1=F(k_1)^{\star}[1+F(k_2)+F(k_3)]-F(k_2)^{\star}F(k_3)\, ,$$ 
 and $f_2$ and $f_3$ can be obtained by cyclicly permuting the momenta in $f_1$, and $F(k)=|\beta_k|^2+\alpha_k^{\star}\beta_k$. The star indicates complex conjugation.  Also, $\Delta_{\mathcal{R}}(k)=\hbar \frac{2\pi^2 }{k^3} \left(\frac{H}{\dot \phi}\right)^2\left(\frac{H}{2\pi}\right)^2$ evaluated at Hubble exit during inflation. 
\end{appendix}

\end{document}